\documentclass[prl,showpacs,showkeys,twocolumn,superscriptaddress,notitlepage,10pt]{revtex4-1}
\usepackage{amsfonts,amsmath,amssymb,graphicx,epstopdf,verbatim,dsfont,color}
\usepackage[english]{babel}

\usepackage[plainpages=false,pdfpagelabels,colorlinks=true,linkcolor=red,urlcolor=blue,citecolor=blue,pdftitle={},pdfauthor={},pdfdisplaydoctitle=true,pdfduplex=DuplexFlipLongEdge]{hyperref}

\def \be{\begin{equation*}}
\def \ee{\end{equation*}}

\begin{document}

\title{Universality in the onset of quantum chaos in many-body systems}

\author{Tyler LeBlond}
\affiliation{Department of Physics, The Pennsylvania State University, University Park, Pennsylvania 16802, USA}
\author{Dries Sels}
\affiliation{Center for Computational Quantum Physics, Flatiron Institute, New York, New York 10010, USA}
\affiliation{Department of Physics,
New York University, New York, New York 10003, USA}
\author{Anatoli Polkovnikov}
\affiliation{Department of Physics, Boston University, Boston, Massachusetts 02215, USA}
\author{Marcos Rigol}
\affiliation{Department of Physics, The Pennsylvania State University, University Park, Pennsylvania 16802, USA}

\begin{abstract}
We show that the onset of quantum chaos at infinite temperature in two many-body one-dimensional lattice models, the perturbed spin-1/2 XXZ and Anderson models, is characterized by universal behavior. Specifically, we show that the onset of quantum chaos is marked by maxima of the typical fidelity susceptibilities that scale with the square of the inverse average level spacing, saturating their upper bound, and that the strength of the integrability- or localization-breaking perturbation at these maxima decreases with increasing system size. We also show that the spectral function below the ``Thouless'' energy (in the quantum-chaotic regime) diverges when approaching those maxima. Our results suggest that, in the thermodynamic limit, arbitrarily small integrability- or localization-breaking perturbations result in quantum chaos in the many-body quantum systems studied here.
\end{abstract}

\maketitle

Quantum chaos and eigenstate thermalization are two intertwined fields that have been the focus of much recent attention in the context of the emergence of statistical mechanics and thermodynamics in isolated quantum systems~\cite{dalessio_kafri_16, deutsch_18, mori_ikeda_18}. Those two fields are built on foundational analytical and computational results~\cite{wigner_55, wigner_57, wigner_58, dyson_62_I, dyson_62_II, dyson_62_III, bohigas_giannoni_84, deutsch_91, srednicki_94, srednicki_99, rigol_dunjko_08}, and they have been recently linked to typicality ideas that date back to von Neumann's work~\cite{vonneumann_10, goldstein_lebowitz_10, rigol_srednicki_12}. When quantum-chaotic systems (which are expected to exhibit eigenstate thermalization) are taken far from equilibrium, few-body operators (observables)  generically equilibrate under unitary dynamics to the predictions of traditional statistical mechanics (they ``thermalize''). This has been verified in experiments with ultracold quantum gases~\cite{trotzky_chen_12, kaufman_tai_16, clos_porras_16, tang_kao_18}. The ``nonthermalizing'' counterparts to quantum-chaotic systems are integrable~\cite{vidmar_rigol_16, essler_fagotti_16, caux_16, calabrese_cardy_16, vasseur_moore_16} and  disorder-localized~\cite{nandkishore_huse_review_15, altman_vosk_review_15, vasseur_moore_16, abanin_altman_19} systems, which have also been probed in experiments with ultracold quantum gases~\cite{kinoshita_wenger_06, gring_kuhnert_12, langen_erne_15, schreiber_hodgman_15, choi_hild_16, tang_kao_18, malvania_zhang_20}. 

In the clean case, a deeper understanding of what happens when quantum-chaotic systems approach integrable points is still needed. In finite systems there is a crossover in which quantum chaos~\cite{rabson_narozhny_04, santos_rigol_10a, rigol_santos_10, santos_rigol_10b, modak_mukerjee_14a, modak_mukerjee_14b, mondaini_fratus_16, mondaini_rigol_17, pandey_claeys_20} and eigenstate thermalization~\cite{rigol_09a, rigol_09b, santos_rigol_10b, mondaini_fratus_16, mondaini_rigol_17} indicators worsen. In the thermodynamic limit one expects quantum chaos and eigenstate thermalization to break down only at the integrable point~\cite{rabson_narozhny_04, santos_rigol_10a, rigol_santos_10, santos_rigol_10b, modak_mukerjee_14a, modak_mukerjee_14b, mondaini_fratus_16, mondaini_rigol_17, pandey_claeys_20} but the time scale for thermalization to diverge approaching that point~\cite{stark_kollar_2013, essler2014quench, bertini2015prethermalization, bertini2016prethermalization, stark_kollar_2013, dalessio_kafri_16, mallayya_rigol_18, mallayya_rigol_19a}. The latter has been seen in recent experiments~\cite{tang_kao_18} and can be understood in the context of Fermi's golden rule~\cite{mallayya_rigol_19a, friedman_2020} and of the scaling of the quantum metric tensor with system size~\cite{pandey_claeys_20}. In the disorder-localized case, localization was argued to be perturbatively stable against weak short-range interactions~\cite{gornyi_mirlin_05, basko_alainer_06} and against strong interactions in one dimension (1D)~\cite{oganesyan_huse_07}. Disorder-induced localization in interacting systems is known as many-body localization and has attracted much theoretical and experimental research in the strongly interacting regime~\cite{nandkishore_huse_review_15, altman_vosk_review_15, vasseur_moore_16, abanin_altman_19}. Recent works have argued against and in favor of the occurrence of many-body localization in that regime in the thermodynamic limit~\cite{suntajs_bonca_19, abanin_bardarson_19, sierant_delande_2020, suntajs_bonca_20, sels_polkovnikov_20, kieferemmanouilidis2020absence}.

We explore the onset of quantum chaos at infinite temperature in perturbed integrable and noninteracting disorder-localized chains, as well as its destruction upon approaching trivial classical limits. One of our goals is to identify universal features and differences between the clean and disordered cases. We compute fidelity susceptibilities $\chi$~\cite{pandey_claeys_20, sierant_maksymov_19}, which are equivalent to the diagonal components of the quantum geometric tensor~\cite{venuti_zanardi_07, kolodrubetz_gritsev_13} or the norm of the adiabatic gauge potential~\cite{pandey_claeys_20}, and spectral functions. Fidelity susceptibilities are commonly used to detect quantum phase transitions~\cite{zanardi_paunkovic_06, venuti_zanardi_07, rigol_shastry_09, kolodrubetz_gritsev_13, kolodrubetz_Sels_17}. We find that the departure from quantum chaos is characterized by a higher sensitivity of eigenstates to perturbations~\cite{pandey_claeys_20, villazon_claeys_20, sels_polkovnikov_20}, which results in maxima of the typical fidelity susceptibility that scale with the square of the inverse level spacing. The shifts in the maxima's positions with system size are consistent with, at infinite temperature in the thermodynamic limit, quantum chaos only failing to occur at the unperturbed integrable, noninteracting disorder-localized, and integrable infinite-interaction (classical) limits.

We study the (clean) extended spin-1/2 XXZ chain:
\begin{equation}\label{eq:cln} 
\hat{H}_\text{cln} = \sum_{i=1}^L \left[ \frac{J}{2} \left( \hat{S}_i^+\hat{S}_{i+1}^- + \text{H.c.} \right) + \Delta \hat{S}_i^z\hat{S}_{i+1}^z + \Delta' \hat{S}_i^z\hat{S}_{i+2}^z \right] ,
\end{equation}
with $J=\sqrt{2}$, $\Delta=(\sqrt{5}+1)/4$, and $\Delta'\in [10^{-4},10^1]$. $\hat{H}_\text{cln}$ is Bethe-ansatz integrable for $\Delta'=0$, and $\hat{H}_\text{cln}/\Delta'$ corresponds to two disconnected Ising chains for $\Delta'=\infty$. We also study the Anderson chain with added nearest-neighbor interactions, which we write in the spin language as
\begin{equation}\label{eq:dsr} 
\hat{H}_\text{dsr} = \sum_{i=1}^L \left[ \frac{J}{2} \left( \hat{S}_i^+\hat{S}_{i+1}^- + \text{H.c.} \right) + h_i \hat{S}_i^z + \Delta \hat{S}_i^z\hat{S}_{i+1}^z \right] ,
\end{equation}
with $J=\sqrt{2}$, $h_i\in[-h,h]$ for $h=(\sqrt{5}+1)/4$, and $\Delta\in [10^{-3},10^1]$~\footnote{For these parameters, the disorder width $(2h\approx 1.62)$ is $\approx 57\%$ of the clean noninteracting electrons bandwidth ($2J\approx2.83$). If one defines the localization length $\xi$ via the single-particle density decay $n_l(x)\sim \exp(-|x-l|/\xi)$, then typical single-particle eigenstates have $\xi \approx 2.6$, while the ones with maximal $\xi$ have $\xi_\text{max}\approx 6$.}. $\hat{H}_\text{dsr}$ is the Anderson model for $\Delta=0$, and $\hat{H}_\text{dsr}/\Delta$ is the Ising chain for $\Delta=\infty$.

To probe the eigenkets $\{|m \rangle\}$ of the models above, we compute the typical fidelity susceptibility $\chi^{}_\text{typ}(O)=\exp(\overline{\ln [\chi_m(O)]})$ (in short, the susceptibility) associated with observable $\hat O$, where
\begin{equation}
\label{eq:chi_n}
 \chi_m(O)=L\sum_{l\neq m}\frac{|\langle m| \hat{O}|l \rangle|^2}{(E_m-E_l)^2}.
\end{equation}
The average $\overline{\ln [\chi_m(O)]}$ is carried out over the central 50\% of eigenstates in the spectrum. We also compute the average spectral function $|f_O(\omega)|^2=\overline{|f^O_m(\omega)|^2}$ over the same 50\% of eigenstates, where  
\begin{equation}
\label{eq:f_n}
|f^O_m(\omega)|^2=L\sum_{l\neq m}|\langle m|\hat{O}|l\rangle|^2 \delta(\omega-\omega_{ml}).
\end{equation}
We replace $\delta(x)\rightarrow\mu/[2\pi(x^2+\mu^2)]$ with $\mu=0.9\, \omega_\text{min}$, where $\omega_\text{min}$ is the minimum level spacing. The factor of $L$ in Eqs.~\eqref{eq:chi_n} and~\eqref{eq:f_n} accounts for the Hilbert-Schmidt norm of our translationally invariant intensive observables.

The specific observables $\hat O$ considered~\cite{suppmat} are the nearest-neighbor ``kinetic'' $\hat K_\text{n}$ and interaction $\hat U_\text{n}$ energies:
\begin{equation}
\hat K_\text{n}=\frac{1}{L}\sum_{i=1}^L \left( \hat{S}_i^+\hat{S}_{i+1}^- + \text{H.c.} \right),\quad \hat U_\text{n}=\frac{1}{L}\sum_{i=1}^L \hat{S}_i^z\hat{S}_{i+1}^z,
\end{equation}
and the next-nearest-neighbor kinetic energy $\hat K_\text{nn}$. As shown recently~\cite{pandey_claeys_20, brenes_goold_20, leblond_rigol_20}, in integrable systems the response of eigenstates to perturbations depends on whether the perturbations do or do not break integrability. If $\hat U_\text{n}$ ($\hat K_\text{nn}$) is added to $\hat{H}_\text{cln}$, integrability is preserved (destroyed), while if $\hat K_\text{n}$ ($\hat U_\text{n}$) is added to $\hat{H}_\text{dsr}$, localization is preserved (destroyed); keeping this in mind, in what follows we show results for $\hat U_\text{n}$ and $\hat K_\text{nn}$ ($\hat K_\text{n}$ and $\hat U_\text{n}$) when studying $\hat{H}_\text{cln}$ ($\hat{H}_\text{dsr}$).

\begin{figure}[!t]
\includegraphics[width=0.985\columnwidth]{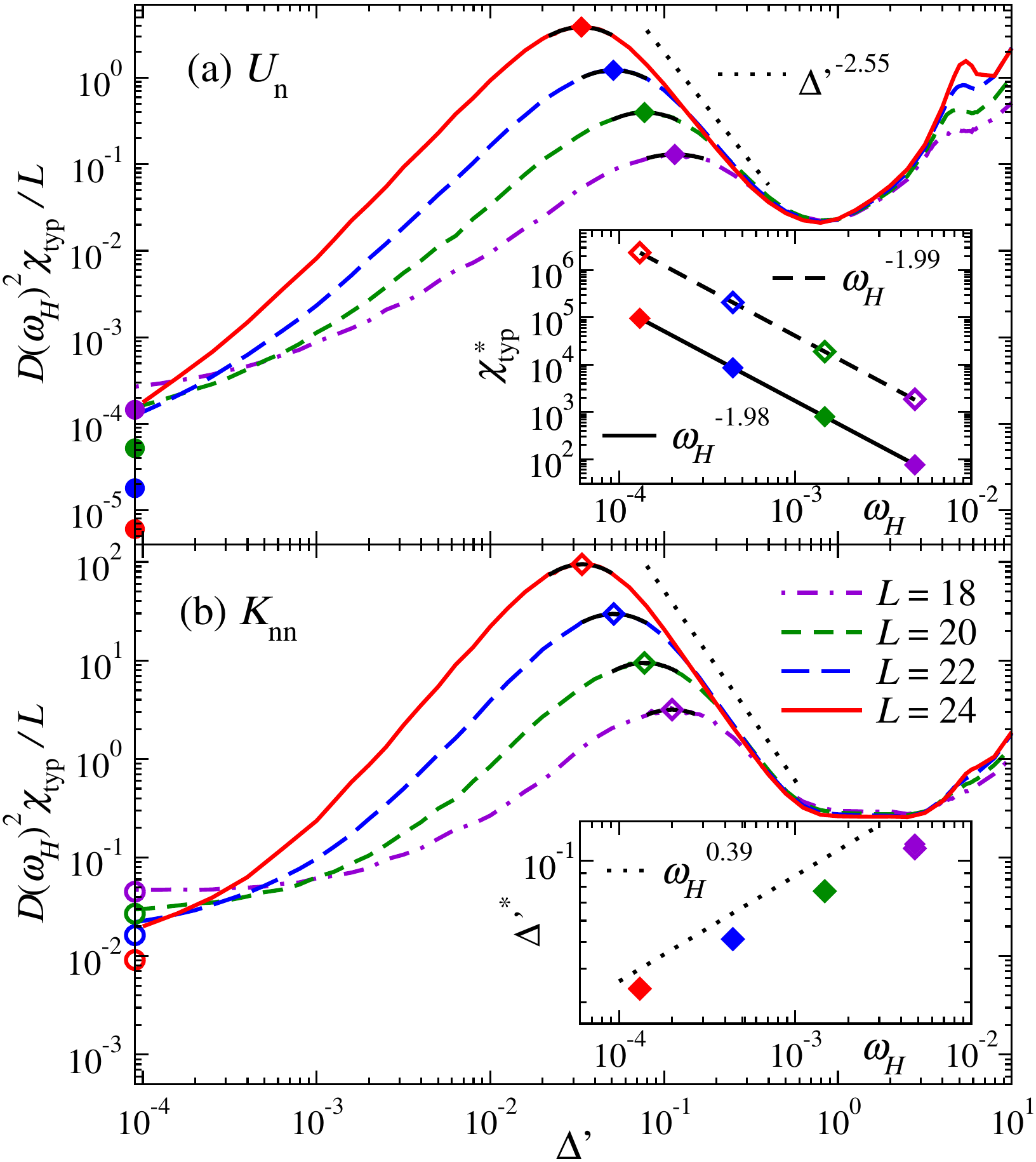}
\caption{\label{fig:1} Typical fidelity susceptibility $\chi^{}_\text{typ}$ (scaled to exhibit collapse in the quantum-chaotic regime) vs the integrability-breaking parameter $\Delta^{\prime}$ for observables $\hat{U}_{\text{n}}$ (a) and $\hat{K}_\text{nn}$ (b) in clean periodic chains. To calculate $\chi^{}_\text{typ}$ and $\omega^{}_H$, we average over the central 50\% of the eigenstates in the even-$Z_2$ sector in each total quasimomentum sector considered. For $L<24$, we report the weighted average over all $k\neq(0,\pi)$ sectors, while for $L=24$ we report results for the $k=\pi/2$ sector. Circles on the $y$-axis show $\chi^{}_\text{typ}$ at the integrable point ($\Delta^{\prime}=0$), and diamonds show the maximal $\chi^*_\text{typ}$ (at $\Delta^{\prime*}=-b/2a$) obtained from polynomial fits $ax^2+bx+c$ (black solid lines about the maxima). The dotted lines on the right of the first peaks are a guide for the eye and depict $\Delta'^{-2.55}$ behavior. Inset in (a): $\chi^*_\text{typ}$ vs $\omega^{}_H$ for both observables, along with the results of power-law fits. Inset in (b): $\Delta^{\prime*}$ vs $\omega_H$ for both observables (the values of $\Delta^{\prime*}$ overlap). The dotted line depicts $\omega^{0.39}_H$ behavior.} 
\end{figure}

In Fig.~\ref{fig:1} we show $\chi^{}_\text{typ}$ vs $\Delta'$ (strength of the integrability-breaking next-nearest-neighbor interaction), for $\hat U_\text{n}$ [Fig.~\ref{fig:1}(a)] and $\hat K_\text{nn}$ [Fig.~\ref{fig:1}(b)]. The susceptibilities are scaled as expected for quantum-chaotic systems, for which $\chi_{\rm typ}\propto LD^{-1}\omega_H^{-2}$ ($\omega_H$ is the mean level spacing and $D$ is the Hilbert space dimension~\footnote{For large systems, $\omega_H\sim \sqrt{L} D^{-1}$, because the effective width of the many-body energy spectrum is $\propto\sqrt{L}$.}) because $|\langle m| \hat{O}|l \rangle|^2 \propto D^{-1}$ for $E_m-E_l\rightarrow\omega_H$~\cite{dalessio_kafri_16, leblond_rigol_20}. For all chain sizes, the scaled susceptibilities exhibit an excellent collapse for about a decade in $\Delta'$ when $\Delta'\sim 1$. The region over which the scaled susceptibilities collapse increases (both towards smaller and towards larger values of $\Delta'$) with increasing system size. This highlights a quantum-chaotic regime that is robust and is increasing its extent with increasing system size,

\begin{figure}[!t]
\includegraphics[width=0.985\columnwidth]{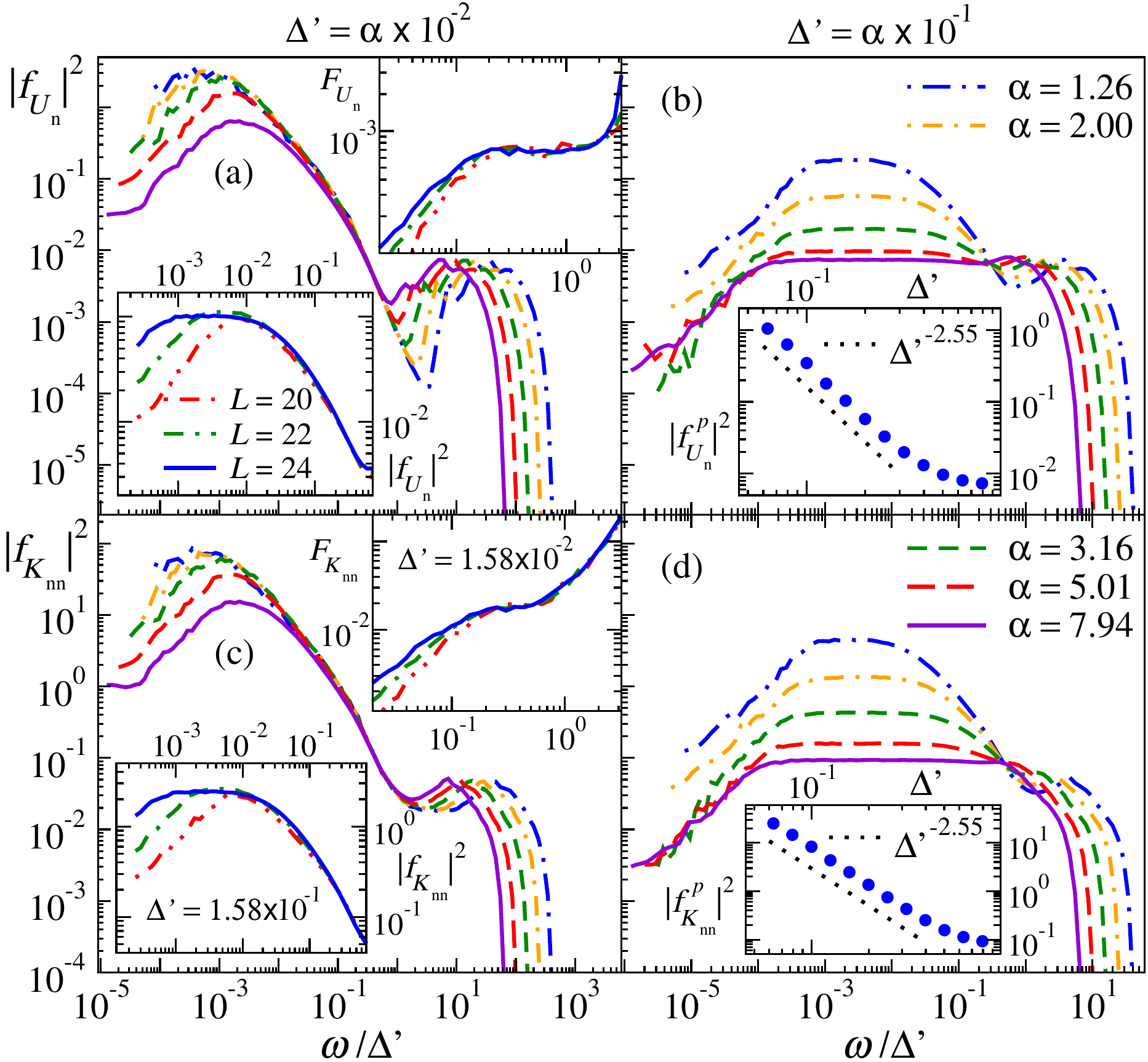}
\caption{\label{fig:2} Spectral functions in clean periodic chains with $L=24$ for observables $\hat{U}_{\text{n}}$ [(a) and (b)] and $\hat{K}_\text{nn}$ [(c) and (d)] over 2 decades of the integrability-breaking parameter $\Delta^{\prime}$ [see labels at the top and legends in (b) and (d)]. In (a) and (c), the top insets show $F_{O} = (\omega/\Delta')^2 |f_O(\omega)|^2$ vs $\omega/\Delta^{\prime}$ at $\Delta^{\prime}=1.58\times10^{-2}$, while the bottom insets show $|f_O(\omega)|^2$ vs $\omega/\Delta^{\prime}$ at $\Delta^{\prime}=1.58\times10^{-1}$, for the three largest chains studied. The insets in (b) and (d) show $|f^p_O(\omega)|^2$ vs $\Delta^{\prime}$, where $|f^p_O(\omega)|^2$ is the value of $|f_O(\omega)|^2$ at the plateaus in the main panels (and for other values of $\Delta'$ for which $|f_O(\omega)|^2$ is not shown). The dotted lines are a guide for the eye and depict $\Delta'^{-2.55}$ behavior. All computations were done as for Fig.~\ref{fig:1}.} 
\end{figure} 

The quantum-chaotic regime in Fig.~\ref{fig:1} is separated from the integrable ones at small and large $\Delta'$ by maxima in $\chi^{}_\text{typ}$~\cite{suppmat}. As a result of the trivial nature of the $\Delta'=\infty$ model, the large-$\Delta'$ maxima are more affected by finite-size effects than the small-$\Delta'$ ones. In what follows we focus on the latter. The inset in Fig.~\ref{fig:1}(a) shows that $\chi^{}_\text{typ}$ at the small $\Delta'$ maxima scales as the square of the inverse average level spacing $\omega_H$. This scaling corresponds to the maximum possible sensitivity of quantum eigenstates to a perturbation~\cite{pandey_claeys_20}. It is exponentially larger, in system size, than expected from random matrix theory. The position of the maxima, $\Delta'^*$, appears to move towards $\Delta'=0$ exponentially fast with increasing system size (notice the near equal shift with increasing $L$ and the log scale in the $\Delta'$ axis). In the inset in Fig.~\ref{fig:1}(b), we plot of $\Delta'^*$ vs $\omega_H$ showing that our numerical results are consistent with $\Delta'^*\propto (\omega_H)^\alpha$, with $\alpha\sim0.39$. We note that our results in Fig.~\ref{fig:1} are robust; $\Delta'^\ast$ and the scaling of $\chi^*_{\rm typ}$ are nearly identical for both observables~\cite{suppmat}.

The susceptibility is related to the spectral function defining the dynamical response of the system~\cite{kolodrubetz_gritsev_13, pandey_claeys_20}. Indeed, it follows from Eqs.~\eqref{eq:chi_n} and~\eqref{eq:f_n} that
\begin{equation}
  \chi_m(O)=\int_{-\infty}^\infty {|f_m^O(\omega)|^2\over \omega^2} d\omega.
\end{equation}

In integrable systems, $|f_O(\omega\to 0)|^2$ vanishes for integrability-preserving perturbations~\cite{pandey_claeys_20, brenes_goold_20, leblond_rigol_20, suppmat}, leading to a polynomial in $L$ scaling of $\overline{\chi_m(O)}$~\cite{pandey_claeys_20}. Typical (integrability breaking) perturbations in contrast have $|f_O(\omega\to 0)|^2=O(1)$~\cite{pandey_claeys_20, brenes_goold_20, leblond_rigol_20, suppmat} resulting in an exponential-in-$L$, $\sim D$, scaling of the susceptibility $\overline{\chi_m(O)}$~\cite{pandey_claeys_20}. As mentioned before, in quantum-chaotic systems $\chi_m(O)\propto L/[D (\omega_H)^2]\sim D$. The faster scaling at the maxima $\chi^*_\text{typ}\propto 1/\omega_H^2\sim D^2$ implies that the spectral function diverges as $|f_O(\omega_H)|^2\sim 1/\omega_H$ around $\Delta'^*$.

Figures~\ref{fig:2}(a) and~\ref{fig:2}(c) show $|f_O(\omega)|^2$ vs $\omega/\Delta'$ for different values of $\Delta'$ about $\Delta'^*$ for $L=24$. The data for both observables collapse at frequencies $\omega/\Delta'\lesssim1$ showing that $|f_O(\omega)|^2 \sim (\Delta'/\omega)^{2}$ in that regime~\footnote{The difference in the extent of the $(\Delta'/\omega)^{2}$ regime is due to the difference in behavior of $|f_O(\omega)|^2$ for integrability-preserving ($\hat U_\text{n}$) vs integrability-breaking ($\hat K_\text{nn}$) operators at $\Delta'=0$, which results in a spectral gap for the former when $\Delta'$ is very small.}. In the top insets, we plot $F_{O}=(\omega/\Delta')^2 |f_O(\omega)|^2$ for different chain sizes when $\Delta'<\Delta'^*$. The plateaus show that the $|f_O(\omega)|^2 \sim (\Delta'/\omega)^{2}$ behavior is robust to changing $L$~\cite{suppmat}. For $\Delta'<\Delta'^*$, the susceptibilities in Figs.~\ref{fig:2}(a) and~\ref{fig:2}(c) also collapse at lower frequencies showing a nontrivial dependence of $\omega/\Delta'$~\cite{suppmat}, but this collapse gradually disappears as $\Delta'$ approaches $\Delta'^*$.

When $\Delta'$ increases beyond $\Delta'^*$ and the system enters into the quantum-chaotic regime [Figs.~\ref{fig:2}(b) and~\ref{fig:2}(d)], a plateau develops in the spectral function at low frequencies~\footnote{For large $L$, a diffusive $1/\sqrt{\omega}$ regime is expected to develop at frequencies above those of the plateau~\cite{dalessio_kafri_16}.}. The formation and growth of the plateau with increasing $L$, at a fixed $\Delta' \gtrsim \Delta'^*$, are illustrated in the bottom insets in Figs.~\ref{fig:2}(a) and~\ref{fig:2}(c). The plateau and the $|f_O(\omega)|^2 \sim (\Delta'/\omega)^{2}$ behavior coexist in the regime in which $\Delta'\gtrsim\Delta'^*$, which is consistent with the occurrence of thermalization with relaxation rates dictated by Fermi's golden rule~\cite{mallayya_rigol_18, suppmat}. In that regime, we find that the spectral function $|f_O(\omega)|^2$ at the plateau, $|f_O^p|^2$, appears to diverge as $(\Delta')^{-\beta}$ with $\beta\sim 2.55$ [see insets in Figs.~\ref{fig:2}(b) and~\ref{fig:2}(d)], consistent with the divergence of $\chi_{\rm typ}$ in Fig.~\ref{fig:1} (see dotted lines in the main panels). Remarkably, it is possible to relate the scaling of $|f_O^p|^2$ with $\Delta'$ with the drift of $\Delta'^\ast$ with $L$: $\Delta'^\ast\sim \omega_H^\alpha$ with $\alpha = 1/\beta \sim 0.39$ [see inset in Fig.~\ref{fig:1}(b)]. 

\begin{figure}[!t]
\includegraphics[width=0.985\columnwidth]{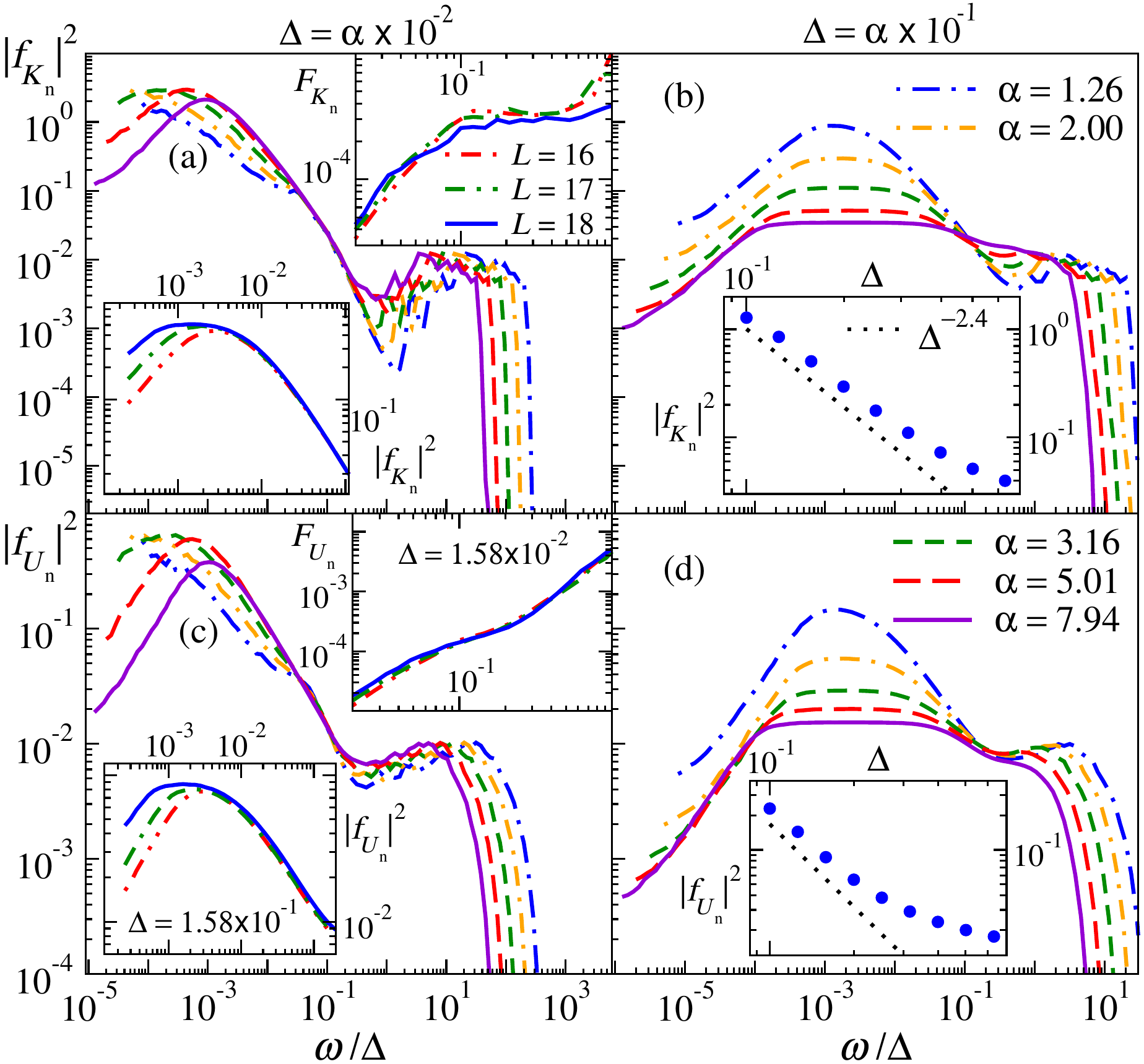}
\caption{\label{fig:3} Spectral functions in disordered periodic chains with $L=18$ for observables $\hat{K}_{\text{n}}$ [(a) and (b)] and $\hat{U}_\text{n}$ [(c) and (d)] over 2 decades of the interaction strength $\Delta$ [see labels at the top, and legends in (b) and (d)]. In (a) and (c), the top insets show $F_{O} = (\omega/\Delta')^2 |f_O(\omega)|^2$ vs $\omega/\Delta^{\prime}$ at $\Delta^{\prime}=1.58\times10^{-2}$, while the bottom insets show $|f_O(\omega)|^2$ vs $\omega/\Delta^{\prime}$ at $\Delta^{\prime}=1.58\times10^{-1}$, for the three largest chains studied. The insets in (b) and (d) show $|f^p_O(\omega)|^2$ vs $\Delta$, where $|f^p_O(\omega)|^2$ is the value of $|f_O(\omega)|^2$ at the plateaus in the main panels (and for other values of $\Delta$ for which $|f_O(\omega)|^2$ is not shown). The dotted lines are a guide for the eye and depict $\Delta'^{-2.4}$ behavior. To calculate $|f_O(\omega)|^2$, we average over the central 50\% of the eigenstates in each chain and then over disorder realizations (200 for $L\leq16$, 100 for $L=17$, and 50 for $L=18$).} 
\end{figure}

We can understand this under the following scenario: Let $|f_O(\omega)|^2 = |f_O^p(\Delta')|^2$ for $\omega < \omega_p(\Delta')$ and $|f_O(\omega)|^2 \propto (\Delta'/\omega)^{\kappa}$ for $\omega > \omega_p(\Delta')$, with $\omega_p(\Delta')$ playing the role of the so-called Thouless energy, and $\kappa>1$. Then from the spectral sum rule, $\int |f_O(\omega)|^2 d\omega=O(1)$, we infer that $\omega_{p}(\Delta')\propto (\Delta')^{\beta}$, with $\beta={\kappa/(\kappa-1)}$, and that $|f_O^p(\Delta')|^2 \propto (\Delta')^{-\beta}$. The maximum of $\chi_{\rm typ}$ then occurs when $\omega_p=\omega_H$, i.e., when the maximum of the spectral function occurs at the Heisenberg scale. This results in $\Delta'^\ast\sim \omega_H^\alpha$ with $\alpha = 1/\beta$, and $\chi_{\rm typ}^*\sim\omega^{-2}_H$. Currently, we do not know the origin of the values of the exponents suggested by our numerical calculations. Given our observation of $|f_O(\omega)|^2 \sim (\Delta'/\omega)^{2}$ behavior for $\Delta'$ below and above $\Delta'^*$, which appears to grow in extent with increasing system size [see top insets in Figs.~\ref{fig:2}(a) and~\ref{fig:2}(c)], two scenarios come to mind: (i) The exponents observed numerically are affected by finite-size effects, and for larger systems than those accessible to us, $\kappa=2$, $\beta=2$, and $\alpha = 1/2$; and (ii) the spectral function develops a power law with an exponent $1<\kappa<2$ before saturating to a constant at low frequencies so that $\beta>2$ and $\alpha < 1/2$.

\begin{figure}[!t]
\includegraphics[width=0.985\columnwidth]{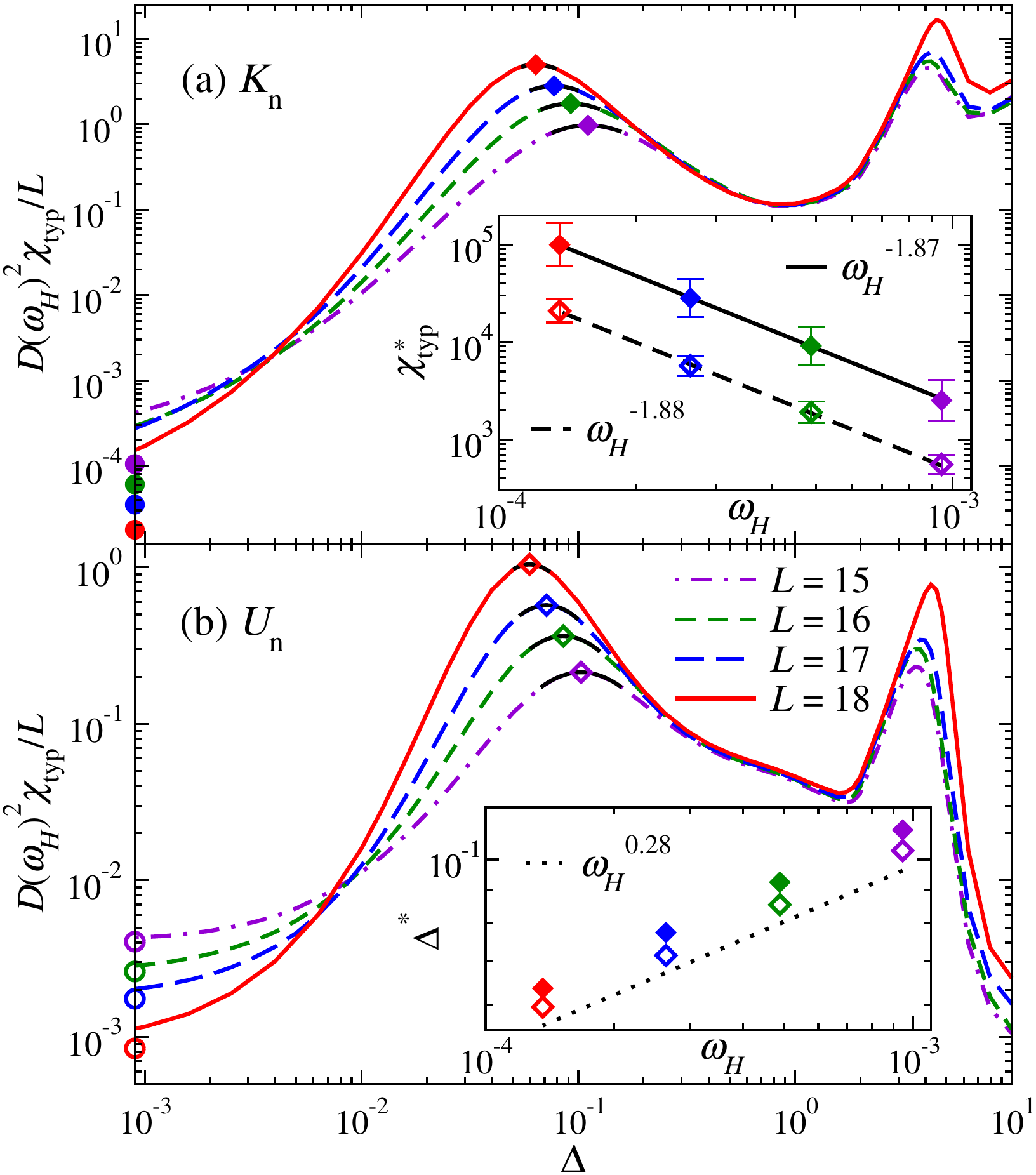}
\caption{\label{fig:4} Typical fidelity susceptibility $\chi^{}_\text{typ}$ (scaled to exhibit collapse in the quantum-chaotic regime) vs the interaction strength $\Delta$ for observables $\hat{K}_{\text{n}}$ (a) and $\hat{U}_\text{n}$ (b) in disordered periodic chains. Circles on the $y$-axis show $\chi^{}_\text{typ}$ at the Anderson-localized point ($\Delta=0$), and diamonds show the maximal $\chi^*_\text{typ}$ (at $\Delta^{*}=-b/2a$) obtained from polynomial fits $ax^2+bx+c$ (black solid lines about the maxima). The inset in (a) shows $\chi^*_\text{typ}$ vs $\omega^{}_H$ for both observables, along with the results of power-law fits. The error bars are the (propagated) standard deviation of the average over disorder realizations (see Ref.~\cite{suppmat} for details) at the value of $\Delta$ (for which we carried out a calculation) that is closest to $\Delta^*$. The inset in (b) shows $\Delta^*$ vs $\omega_H$ for both observables. The dotted line depicts $\omega_H^{0.28}$ behavior. All computations were done as for Fig.~\ref{fig:3}.} 
\end{figure}

In Fig.~\ref{fig:3}, we show results for the spectral function of disordered chains in the presence of nearest-neighbor interactions. The corresponding typical fidelity susceptibilities are shown in Fig.~\ref{fig:4}. The results in Figs.~\ref{fig:3} and~\ref{fig:4} are similar to those in Figs.~\ref{fig:2} and~\ref{fig:1}, respectively. The similarity is remarkable considering that the unperturbed models in both cases are strikingly different, the disordered one being a noninteracting localized model and the clean one being an interacting integrable one. The slight differences between the results in Figs.~\ref{fig:3} and~\ref{fig:2} include a narrower $|f_O(\omega)|^2 \sim (\Delta'/\omega)^{2}$ regime in Figs.~\ref{fig:3}(a) and~\ref{fig:3}(c) as compared with Figs.~\ref{fig:2}(a) and~\ref{fig:2}(c), and a narrower regime in which $|f_O^p|^2$ is consistent with a power law scaling with $\Delta$ in Fig.~\ref{fig:3}(d). Related to the latter, in the inset in Fig.~\ref{fig:4}(b) the dynamical range for $\Delta^*$ vs $\omega_H$ is smaller than in the inset in Fig.~\ref{fig:1}(b). Consequently, and also keeping in mind that in Fig.~\ref{fig:4} we plot typical fidelity susceptibilities while in Fig.~\ref{fig:3} we plot raw averages of the spectral functions, we cannot establish a relationship between the scaling of $|f_O^p|^2$ with $\Delta'$ and the drift of $\Delta'^\ast$ with $L$ as we did for the clean case. That said, all those differences are consistent with stronger finite-size effects, and fluctuations associated with the disorder average, in the disordered systems. For the latter, the largest chains studied have $L=18$ vs the $L=24$ chains considered for clean systems.

In summary, our results suggest that the onset of quantum chaos at infinite temperature in the models studied, as well as its destruction when approaching classical limits for very strong interactions, is characterized by universal behavior. We focused our analysis on the onset of quantum chaos as finite-size effects (and fluctuations associated with disorder averages) are smaller. The main universal feature identified is the divergence of the typical fidelity susceptibilities as $\omega_H^{-2}$ when entering (exiting) the quantum-chaotic regime and the associated divergence of the spectral functions below the Thouless energy. The latter is potentially universal and diverges as $\epsilon^{-\beta}$ ($\epsilon$ being the strength of either the integrability-breaking or localization-breaking perturbation) in the quantum-chaotic regime. Also potentially universal is the shift of the position $\epsilon^*$ of the maximum of the fidelity susceptibilities as $\epsilon^*\sim\omega_H^\alpha$, as well as the relation $\alpha=1/\beta$ between the exponents. We note that $\epsilon^*\sim\omega_H^\alpha$ supports the expectation that in clean systems in the thermodynamic limit, quantum chaos and eigenstate thermalization break down only at the integrable point~\cite{rabson_narozhny_04, santos_rigol_10a, rigol_santos_10, santos_rigol_10b, modak_mukerjee_14a, modak_mukerjee_14b, mondaini_fratus_16, mondaini_rigol_17, pandey_claeys_20}, and it suggests that at infinite temperature the 1D Anderson insulator (for the parameters considered here) is unstable against adding interactions. An interesting open question is whether this relates to recent findings that many-body localization is unstable against the insertion of thermal ``bubbles'' if disorder is not strong enough~\cite{deroeck17, crowley2020}.

Much still needs to be explored, such as what happens at finite temperatures and when one changes the parameters of the unperturbed Hamiltonians [which we selected to be $O(1)$ to minimize finite-size effects]. In the disordered case, two parameter regimes to be explored are the strong disorder and strong interaction regimes. The contrast between the small $\Delta$ and large $\Delta$ peaks in the fidelity susceptibilities in Fig.~\ref{fig:4} suggests that obtaining meaningful scalings using full exact diagonalization in those regimes will be computationally very challenging. We note that the results reported in this Research Letter required about 1\,000\,000 CPU hours of calculations.

We acknowledge discussions with E.~Altman, A.~Dymarsky, S.~Gopalakrishnan, D.~Huse, M.~Pandey, and L.~Vidmar. This work was supported by the National Science Foundation under Grants No.~PHY-2012145 (T.L. and M.R.), No.~DMR-1813499 (A.P.), and No.~DMR-2103658 (A.P.), and by the AFOSR under Grants No.~FA9550-16-1-0334 (A.P.), No.~FA9550-21-1-0342 (A.P.), and No.~FA9550-21-1-0236 (D.S.). The computations were carried out in the Roar supercomputer in the Institute for Computational and Data Sciences (ICDS) at Penn State. The Flatiron Institute is a division of the Simons Foundation. 

\bibliographystyle{biblev1}
\bibliography{refs}

\newpage
\phantom{a}
\newpage
\setcounter{figure}{0}
\setcounter{equation}{0}
\setcounter{table}{0}

\renewcommand{\thetable}{S\arabic{table}}
\renewcommand{\thefigure}{S\arabic{figure}}
\renewcommand{\theequation}{S\arabic{equation}}
\renewcommand{\thepage}{S\arabic{page}}

\renewcommand{\thesection}{S\arabic{section}}

\onecolumngrid

\begin{center}

\setcounter{page}{1}

{\large \bf Supplemental Material:\\
Universality in the Onset of Quantum Chaos in Many-Body Systems}\\

\vspace{0.3cm}

Tyler LeBlond$^{1}$, Dries Sels$^{2,3}$, Anatoli Polkovnikov$^4$ and Marcos Rigol$^{1}$\\
$^1${\it Department of Physics, The Pennsylvania State University, University Park, Pennsylvania 16802, USA} \\
$^2${\it Center for Computational Quantum Physics, Flatiron Institute, New York, New York 10010, USA} \\ 
$^3${\it Department of Physics, New York University, New York, New York 10003, USA} \\
$^4${\it Department of Physics, Boston University, Boston, Massachusetts 02215, USA}

\end{center}

\vspace{0.6cm}

\twocolumngrid

\label{pagesupp}

\section{Additional numerical results for clean systems}

\begin{figure}[!b]
\includegraphics[width=0.985\columnwidth]{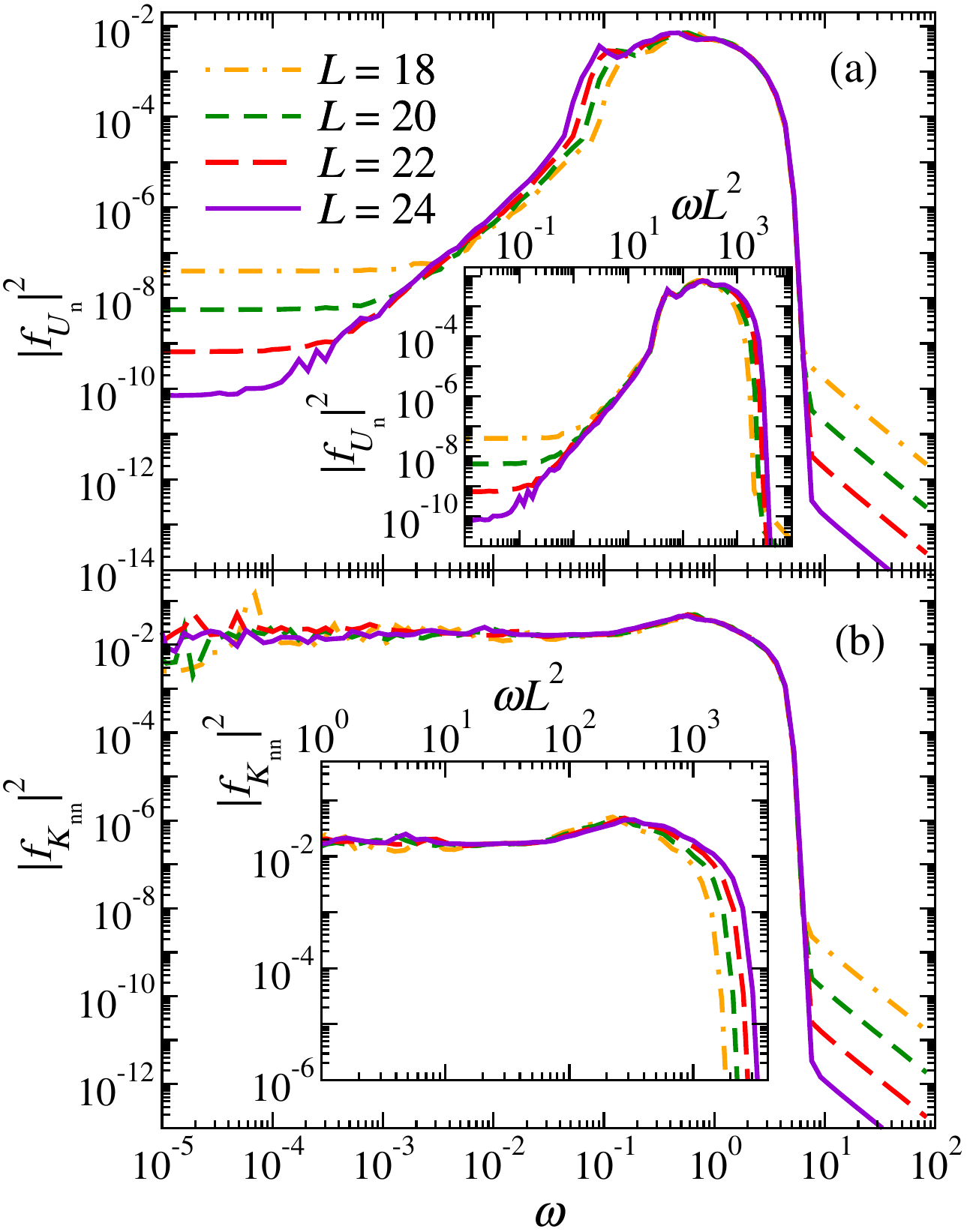}
\caption{\label{sm_fig:1} Spectral functions in clean periodic chains for observables $\hat{U}_{\text{n}}$ (a) and $\hat{K}_\text{nn}$ (b) at the integrable point $\Delta^{\prime}=0$. In our calculations we average over the central 50\% of the eigenstates in the even-$Z_2$ sector in each total quasimomentum sector considered. For $L<24$, we report the weighted average over all $k\neq(0,\pi)$ sectors, while for $L=24$ we report results for the $k=\pi/2$ sector.} 
\end{figure} 

In Fig.~\ref{sm_fig:1} we plot $|f_O(\omega)|^2$ vs $\omega$, for observables $\hat U_\text{n}$ [Fig.~\ref{sm_fig:1}(a)] and $\hat K_\text{nn}$ [Fig.~\ref{sm_fig:1}(b)], at the integrable point for chains with $L=18$ through $L=24$. These results show that for the integrability preserving operator ($\hat U_\text{n}$) the spectral function plateaus (as $\omega$ approaches $\omega_H$) at a value that decreases exponentially with $L$, while for the integrability breaking operator ($\hat K_\text{nn}$) the spectral function plateaus at an $O(1)$ value, as found in Ref.~\cite{pandey_claeys_20}. In the insets in Fig.~\ref{sm_fig:1}, the low-frequency data collapse for different values of $L$ show that the spectral function is a function of $\omega L^2$ at low frequencies, as found in Ref.~\cite{leblond_rigol_20}.

\begin{figure}[!b]
\includegraphics[width=0.985\columnwidth]{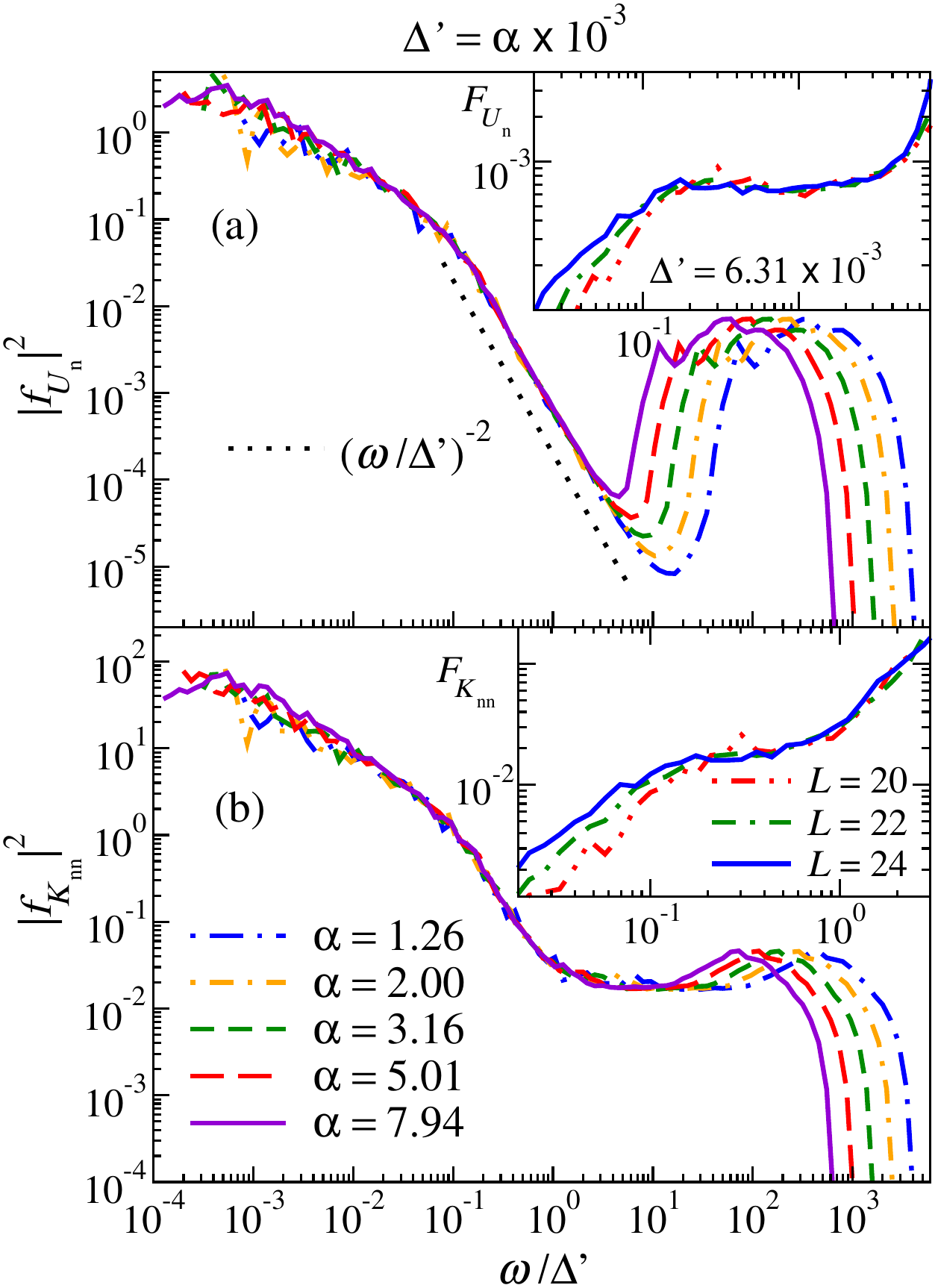}
\caption{\label{sm_fig:2} Spectral functions in clean periodic chains with $L=24$ for observables $\hat{U}_{\text{n}}$ (a) and $\hat{K}_\text{nn}$ (b) over one decade of the integrability-breaking parameter $\Delta^{\prime}$ [see label at the top and legends in (b)]. The insets show $F_{O} = (\omega/\Delta')^2 |f_O(\omega)|^2$ vs $\omega/\Delta^{\prime}$ at $\Delta^{\prime}=6.31\times10^{-3}$ for the three largest chains studied. All computations were done as for Fig.~\ref{sm_fig:1}.} 
\end{figure}

In Fig.~\ref{sm_fig:2} we plot $|f_O(\omega)|^2$ vs $\omega/\Delta'$, for observables $\hat U_\text{n}$ [Fig.~\ref{sm_fig:2}(a)] and $\hat K_\text{nn}$ [Fig.~\ref{sm_fig:2}(b)], when $10^{-3}\leq\Delta'\leq10^{-2}$ for chains with $L=24$. These results are the lower $\Delta'$ precursors of the results shown in Figs.~2(a) and~2(c) in the main text. The plateaus in the insets make apparent the robust with increasing system size $(\Delta'/\omega)^{2}$ behavior in the spectral functions for $\Delta'<\Delta'^*$.

\begin{figure}[!t]
\includegraphics[width=0.985\columnwidth]{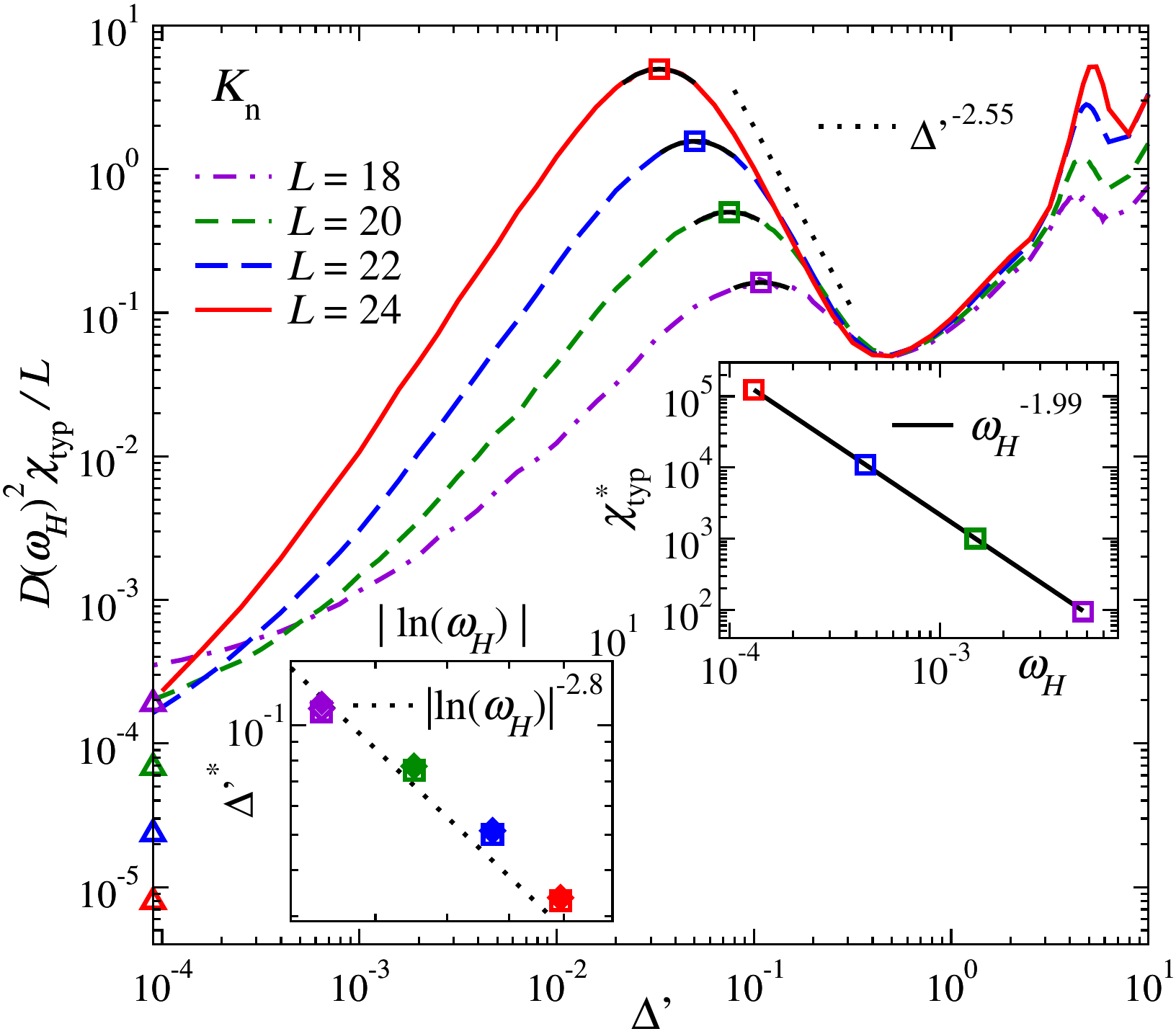}
\caption{\label{sm_fig:3} Typical fidelity susceptibility $\chi^{}_\text{typ}$ (scaled to exhibit collapse in the quantum-chaotic regime) vs the integrability-breaking parameter $\Delta^{\prime}$ for $\hat{K}_\text{n}$. Triangles on the y-axis show $\chi^{}_\text{typ}$ at the integrable point ($\Delta^{\prime}=0$), and squares show the maximal $\chi^*_\text{typ}$ (at $\Delta^{\prime*}=-b/2a$) obtained from polynomial fits $ax^2+bx+c$ (black solid lines about the maxima). The straight dotted line on the right of the first peaks depicts $\Delta'^{-2.55}$ behavior. Right inset: $\chi^*_\text{typ}$ vs $\omega^{}_H$, along with the result of a power-law fit. Left inset: $\Delta'^*$ vs $|\ln (\omega_H)|$ for $\hat{K}_\text{n}$, $\hat U_\text{n}$, and $\hat K_\text{nn}$ (the values of $\Delta'^*$ for the three observables overlap). The dotted line depicts $|\ln (\omega_H)|^{-2.8}$ behavior. All computations were done as for Figs.~\ref{sm_fig:1} and~\ref{sm_fig:2}.} 
\end{figure}

In Fig.~\ref{sm_fig:3}, we plot the (scaled) typical fidelity susceptibility for the nearest-neighbor kinetic energy $\hat K_\text{n}$ in clean systems. These results are the equivalent of the ones reported in Fig.~1 of the main text for $\hat U_\text{n}$ and $\hat K_\text{nn}$. The results in Fig.~\ref{sm_fig:3} are very similar to those in Fig.~1, and are most similar to the ones reported in Fig.~1(a). This is expected as $\hat K_\text{n}$ is an operator that if added as a perturbation to the spin-1/2 XXZ Hamiltonian preserves integrability, like $\hat U_\text{n}$. Hence, the scaling of the typical fidelity susceptibilities is the same for $\hat K_\text{n}$ and $\hat U_\text{n}$ at the unperturbed integrable point. 

The right inset in Fig.~\ref{sm_fig:3} shows that the scaling of $\chi^*_\text{typ}$ for $\hat K_\text{n}$ is the same as for $\hat U_\text{n}$ and $\hat K_\text{nn}$ in the main text. The left inset in Fig.~\ref{sm_fig:3} shows that $\Delta'^*$ is less consistent with a polynomial scaling in $\ln (\omega_H)$ (polynomial in $L$) than with a polynomial scaling in $\omega_H$ (exponential in $L$) as shown in the main text. If it were to be a polynomial scaling in $\ln (\omega_H)$ then the power would be large ($\Delta'^*\sim |\ln (\omega_H)|^{-2.8}$).

\begin{figure}[!t]
\includegraphics[width=0.985\columnwidth]{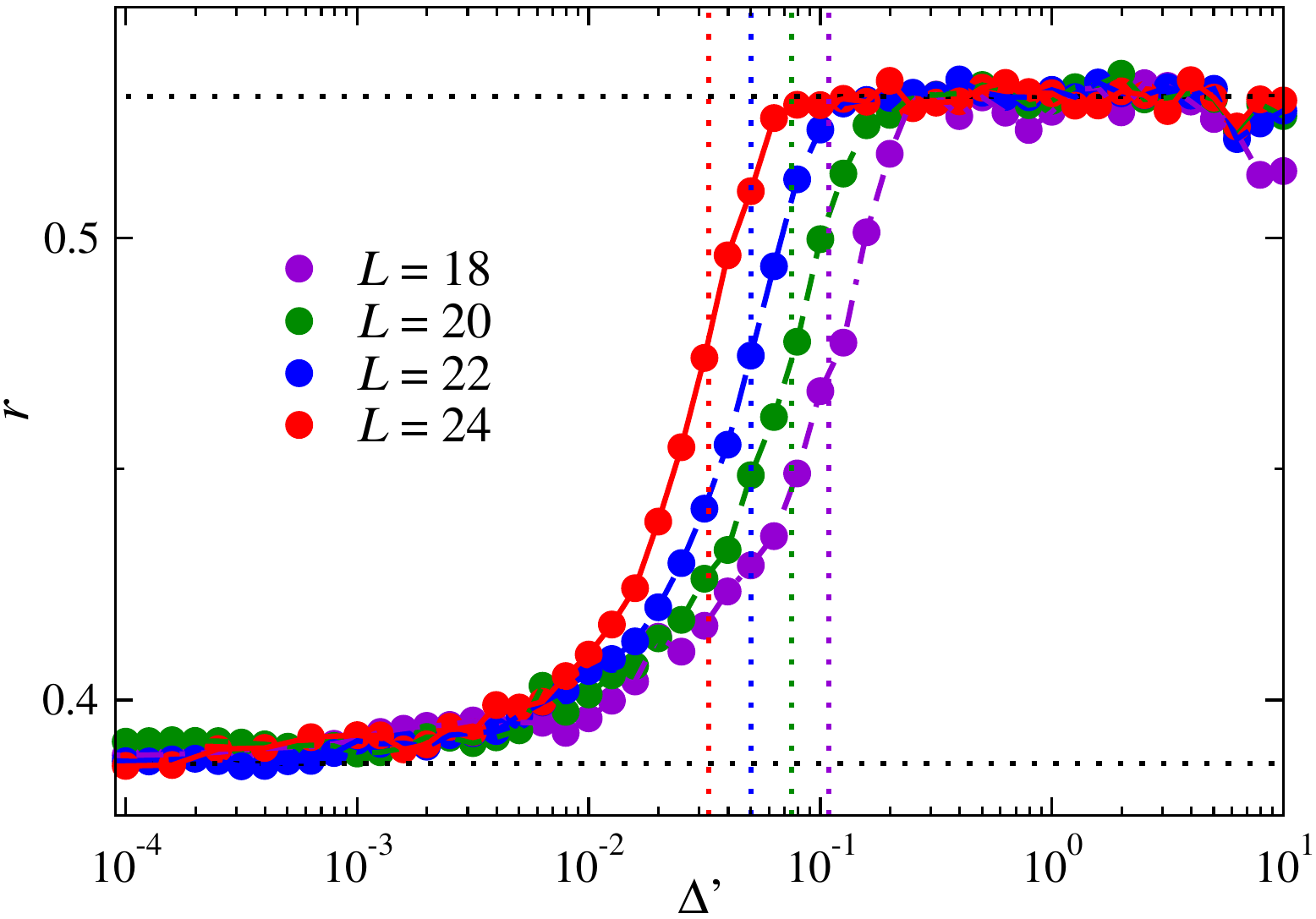}
\caption{\label{sm_fig:levclean} Average ratio of level spacings $r$ vs the integrability-breaking parameter $\Delta^{\prime}$. The horizontal dotted lines show the predictions for the Gaussian orthogonal ensemble $r_\text{GOE}\approx0.5307$ and for the Poisson distribution $r_\text{P}\approx0.3863$~\cite{atas_bogomolny_13}. The vertical dotted lines show the positions of the maxima of the typical fidelity susceptibility identified in the main panel in Fig.~\ref{sm_fig:3}. All averages were done over the same part of the spectrum as for Figs.~\ref{sm_fig:1}--\ref{sm_fig:3}.} 
\end{figure}

In Fig.~\ref{sm_fig:levclean}, we plot the average ratio of level spacings $r=\overline{r_m}$ for the central 50\% of the energy eigenvalues in the same quasimomentum sectors used for the previous figures. The ratio of level spacings for eigenstate $|m\rangle$ is defined as $r_m = \text{min}\{\delta_m, \delta_{m-1} \}/ \text{max}\{\delta_m , \delta_{m-1}\}$, where $\delta_m = E_{m+1} - E_m$~\cite{oganesyan_huse_07}. Two things to note are: (i) With increasing system size the departure from the prediction of the Gaussian orthogonal ensemble, $r_\text{GOE}\approx0.5307$, shifts towards smaller values of the integrability breaking parameter. This is similar to the behavior observed in earlier works for other quantum chaos indicators related to the statistical properties of the level spacings, e.g., the position of the maximum of the level spacing distribution in Fig.~4 in Ref.~\cite{santos_rigol_10a} and the parameter $\beta$ obtained fitting the level spacing distribution to a Brody distribution in Fig.~1 in Ref.~\cite{rigol_santos_10}, as well as to the properties of the energy eigenstates, e.g., in Fig.~2 in Ref.~\cite{mondaini_rigol_17}. The results in Fig.~\ref{sm_fig:levclean} show that the shift in the departure of $r$ from the prediction of the Gaussian orthogonal ensemble resembles the shift of the maxima of the typical fidelity susceptibility discussed before. (ii) It is not possible to identify a clear crossing in the curves of $r$ vs $\Delta'$ for different systems sizes.

\section{Additional numerical results for disordered systems}

\begin{figure}[!t]
\centering
\includegraphics[width=0.985\columnwidth]{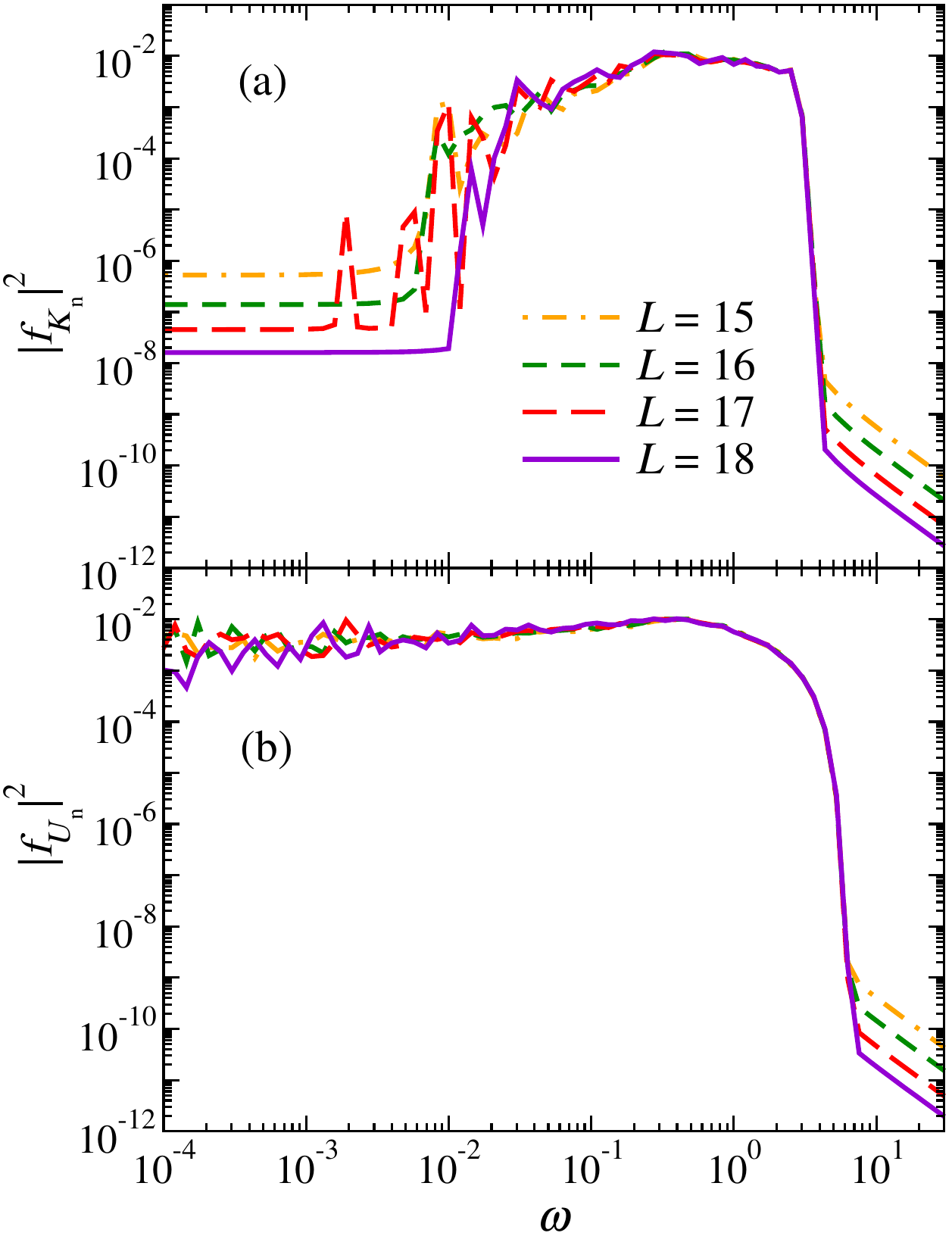}
\caption{\label{sm_fig:4} Spectral functions in disordered periodic chains for observables $\hat{K}_\text{n}$ (a) and $\hat{U}_\text{n}$ (b) at the noninteracting point $\Delta=0$. To calculate $|f_O(\omega)|^2$, we average over the central 50\% of the eigenstates in each chain, and then over disorder realizations (200 for $L\leq16$, 100 for $L=17$, and 50 for $L=18$).} 
\end{figure} 

In Fig.~\ref{sm_fig:4} we plot $|f_O(\omega)|^2$ vs $\omega$, for observables $\hat K_\text{n}$ [Fig.~\ref{sm_fig:1}(a)] and $\hat U_\text{n}$ [Fig.~\ref{sm_fig:1}(b)], at the noninteracting point for disordered chains with $L=15$ through $L=18$. These results show that for the localization preserving operator ($\hat K_\text{n}$) the spectral function plateaus (as $\omega\rightarrow\omega_H$) at a value that decreases exponentially with $L$ while for the localization breaking operator ($\hat U_\text{n}$) the spectral function plateaus at an $O(1)$ value, as in Fig.~\ref{sm_fig:1} for integrability preserving and breaking operators, respectively.

\begin{figure}[!t]
\includegraphics[width=0.985\columnwidth]{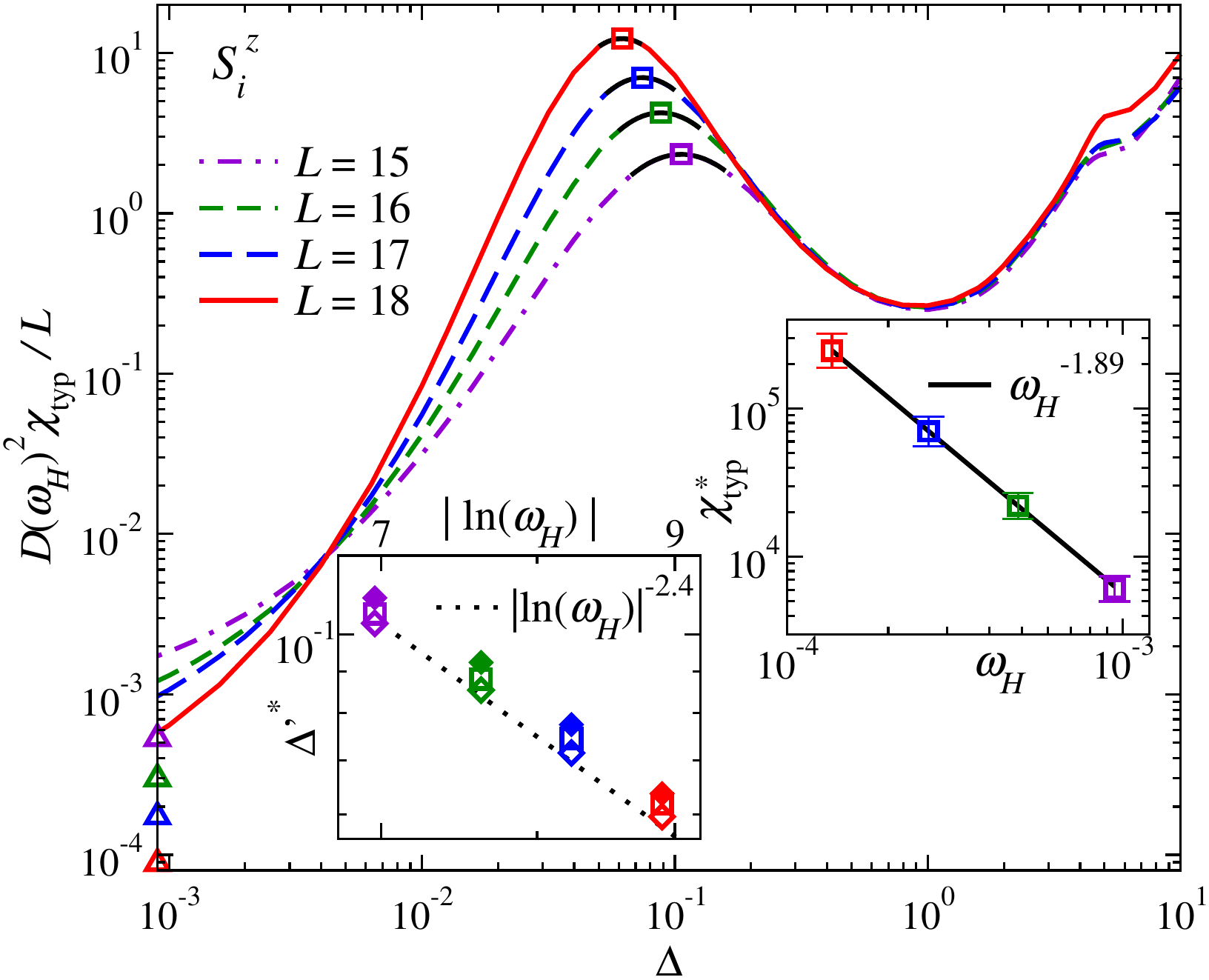}
\caption{\label{sm_fig:5} Typical fidelity susceptibility $\chi^{}_\text{typ}$ (scaled to exhibit collapse in the quantum-chaotic regime) vs the interaction strength $\Delta$ for $\hat{S}^{z}_{i}$ in disordered periodic chains. Triangles on the y-axis show $\chi^{}_\text{typ}$ at the Anderson-localized point ($\Delta=0$), and squares show the maximal $\chi^*_\text{typ}$ (at $\Delta^{*}=-b/2a$) obtained from polynomial fits $ax^2+bx+c$ (black solid lines about the maxima). To calculate $\chi^{}_\text{typ}$, we average over the central 50\% of the eigenstates in each chain, over sites, and over disorder realizations (all sites and 200 disorder realizations for $L\leq16$, 8 sites and 100 disorder realizations for $L=17$, and 3 sites and 50 disorder realizations for $L=18$). Right inset: $\chi^*_\text{typ}$ vs $\omega^{}_H$, along with the result of a power-law fit. The error bars are the (propagated) standard deviation of the average over disorder realizations at the value of $\Delta$ (for which we carried out a calculation) closest to $\Delta^*$. Left inset: $\Delta'^*$ vs $|\ln (\omega_H)|$ for $\hat{S}^{z}_{i}$, $\hat K_\text{n}$, and $\hat U_\text{n}$. The dotted line depicts $|\ln (\omega_H)|^{-2.4}$ behavior. } 
\end{figure}

In Fig.~\ref{sm_fig:5}, we plot the (scaled) typical fidelity susceptibility for the local $\hat{S}^{z}_{i}$ operator in disordered systems. [Note that, since $\hat{S}^{z}_{i}$ has an $O(1)$ norm, $\chi_n$ for this operator lacks the factor of $L$ present in Eq.~(3) in the main text.] These results are the equivalent of the ones reported in Fig.~4 of the main text for $\hat K_\text{n}$ and $\hat U_\text{n}$. The results in Fig.~\ref{sm_fig:5} are very similar to those in Fig.~4, and are most similar to the ones reported in Fig.~4(a). This is expected as $\hat{S}^{z}_{i}$ is an operator that if added as a perturbation to the Anderson chain preserves localization, like $\hat K_\text{n}$.

The right inset in Fig.~\ref{sm_fig:5} shows that the scaling of $\chi^*_\text{typ}$ for $\hat{S}^{z}_{i}$ is the same as for $\hat K_\text{n}$ and $\hat U_\text{n}$ in the main text. The left inset in Fig.~\ref{sm_fig:3} shows that $\Delta'^*$ is less consistent with a polynomial scaling in $\ln (\omega_H)$ (polynomial in $L$) than with a polynomial scaling in $\omega_H$ (exponential in $L$) as shown in the main text. If it were to be a polynomial scaling in $\ln (\omega_H)$ then the power would be large ($\Delta'^*\sim |\ln (\omega_H)|^{-2.4}$).

\begin{figure}[!t]
\includegraphics[width=0.985\columnwidth]{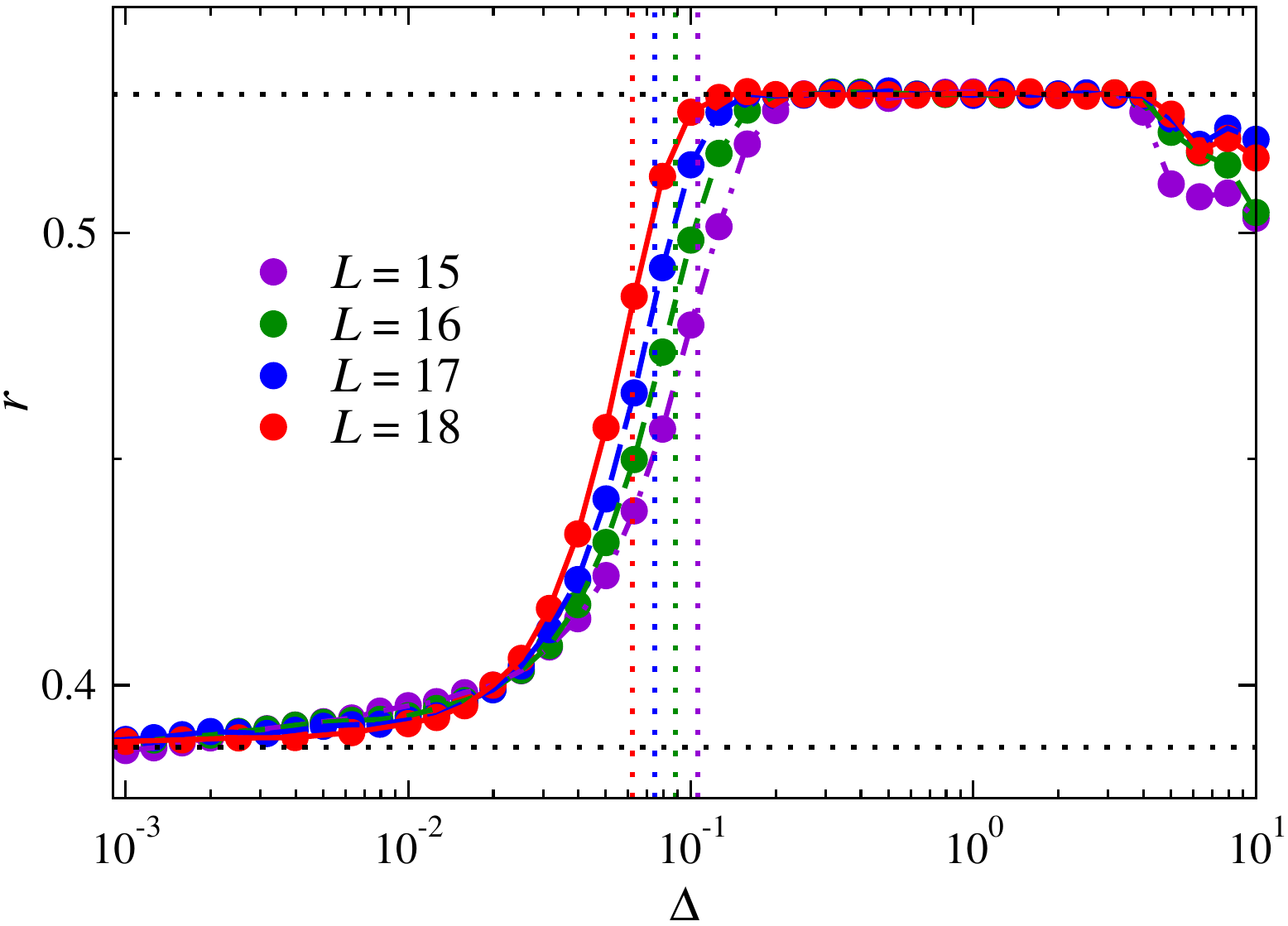}
\vspace{-0.1cm}
\caption{\label{sm_fig:levdisr} Average ratio of level spacings $r$ vs the interaction strength $\Delta$ in disordered periodic chains. The horizontal dotted lines show the predictions for the Gaussian orthogonal ensemble $r_\text{GOE}\approx0.5307$ and for the Poisson distribution $r_\text{P} \approx 0.3863$~\cite{atas_bogomolny_13}. The vertical dotted lines show the positions of the maxima of the typical fidelity susceptibility identified in the main panel in Fig.~\ref{sm_fig:5}. All averages were done over the same part of the spectrum as for Figs.~\ref{sm_fig:4} and~\ref{sm_fig:5}.} 
\end{figure}

In Fig.~\ref{sm_fig:levdisr}, we plot the average ratio of level spacings $r=\overline{r_n}$ in the disordered systems. The results are qualitatively similar to those in Fig.~\ref{sm_fig:levclean}. We do note that while the results in Fig.~\ref{sm_fig:levdisr} for $r\sim r_\text{p}$ are qualitatively similar to those in Fig.~\ref{sm_fig:levclean}, for disordered systems the fluctuations in the data are smaller (because of the disorder average) and the curves for the four system sizes shown cross around $\Delta\approx0.02$. Given the shift in the maxima of the typical fidelity susceptibilities with increasing system size discussed before, and the shift in the departure of $r$ from $r_\text{GOE}$ seen in Fig.~\ref{sm_fig:levdisr}, we expect crossings as well as the entire $r\sim r_\text{p}$ part of the curves to shift towards $\Delta=0$ with increasing system size.

\section{Standard deviation of the disorder averages}

\begin{figure}[!b]
\includegraphics[width=0.985\columnwidth]{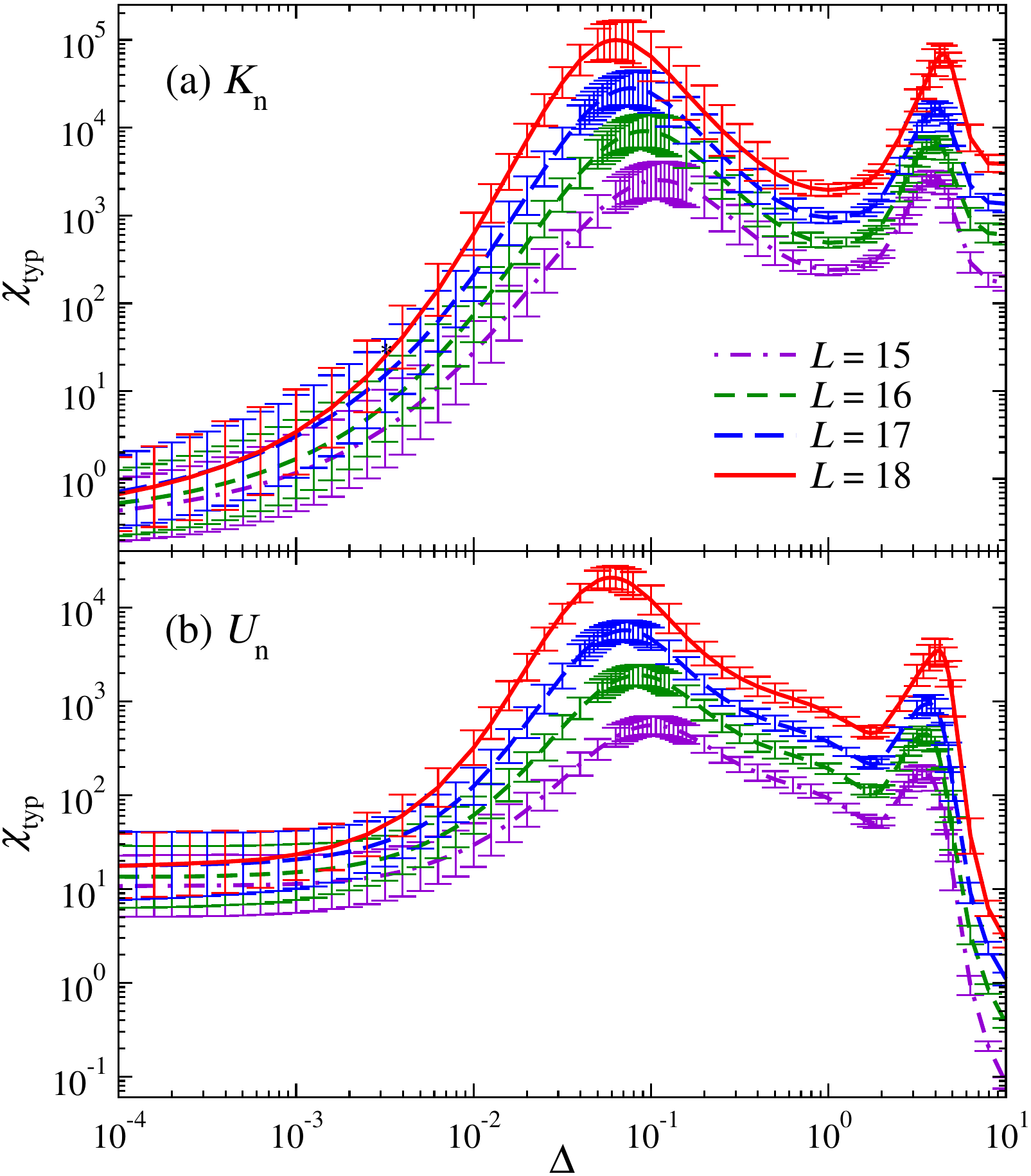}
\vspace{-0.1cm}
\caption{\label{fig:4_error} $\chi_\text{typ}$ corresponding to the main panels in Fig.~4 in the main text but including error bars, which are the (propagated) standard deviation of the average over disorder realizations at each value of $\Delta$. See text for details. The values of $\Delta$ were selected to be equally spaced in a logarithmic scale, and extra finer grids were used about the maxima in the typical fidelity susceptibilities.} 
\end{figure}

\begin{figure}[!t]
\includegraphics[width=0.985\columnwidth]{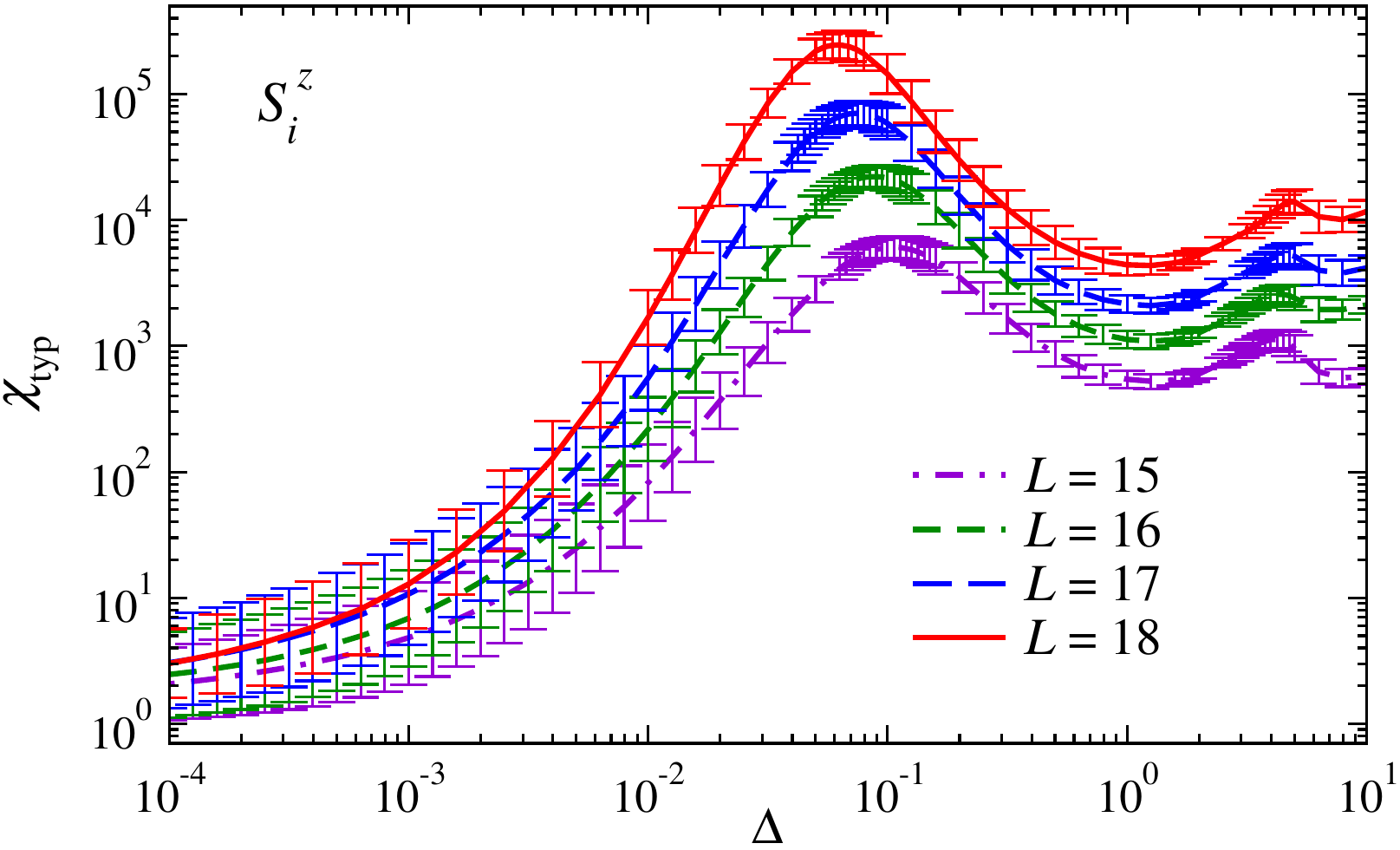}
\vspace{-0.1cm}
\caption{\label{sm_fig:5_error} $\chi_\text{typ}$ corresponding to the main panel in Fig.~\ref{sm_fig:5} but including error bars, which are the (propagated) standard deviation of the average over disorder realizations at each value of $\Delta$ studied. See text for details. The values of $\Delta$ were selected to be equally spaced in a logarithmic scale, and extra finer grids were used about the maxima in the typical fidelity susceptibilities.} 
\end{figure}

In Fig.~\ref{fig:4_error} we show $\chi_\text{typ}$ corresponding with the results in Fig.~4 in the main text along with the (propagated) standard deviation of the average over disorder realizations. Namely, we compute the standard deviation of the average of the logarithms and then exponentiate them as in $\chi_\text{typ}^{\pm} = \exp{(\overline{\ln{\chi_n}}\pm \sigma)}$, where $\sigma = \sqrt{\langle\overline{\ln{\chi_n}}^2\rangle - \langle\overline{\ln{\chi_n}}\rangle^2}$ and $\langle...\rangle$ is an average over disorder realizations. In Fig.~\ref{sm_fig:5_error}, we show $\chi_\text{typ}$ corresponding to the results shown in the main panel in Fig.~\ref{sm_fig:5} along with the (propagated) standard deviation of the average over disorder realizations. For all observables, the errors are smallest in the regime in which the system exhibits eigenstate thermalization and largest for $\Delta\lesssim 10^{-2}$.

\section{$(\epsilon/\omega)^{2}$ perturbative scaling and Fermi's golden rule}

We use perturbation theory to analyze the spectral function of a weakly perturbed integrable Hamiltonian. Let us assume that the Hamiltonian can be written as
\begin{equation}
    \hat H=\hat H_0+\epsilon \hat V,
\end{equation}
where $\hat H_0$ is an integrable Hamiltonian and $\hat V$ is the integrability breaking perturbation. For simplicity, we assume that the diagonal matrix elements of $\hat V$ in the eigenstates of $\hat H_0$ vanish. They produce shifts in the perturbed eigenenergies without affecting the eigenstates, and can be absorbed in the definition of $\hat H_0$.

Let us compute the leading perturbative correction to the magnitude of the matrix elements $|\langle m | \hat O |l\rangle|^2$ of an arbitrary operator $\hat O$ in the perturbed Hamiltonian eigenstates $\{|m\rangle\}$. Expanding 
\begin{equation}
O_{ml}\equiv \langle m| \hat O|l\rangle=O_{ml}^{(0)}+\epsilon\, O_{ml}^{(1)}+\dots\,,
\end{equation}
and applying standard perturbation theory, one finds
\begin{eqnarray}\label{eq:mateleper}
O_{ml}^{(1)}&=&-{O_{mm}-O_{ll}\over E^{(0)}_m-E^{(0)}_l} V_{ml}\nonumber \\ &&
+\sum_{k\neq m,l} {O_{mk} V_{kl}\over E^{(0)}_l-E^{(0)}_k}+{V_{mk} O_{kl}\over E^{(0)}_m-E^{(0)}_k},
\end{eqnarray}
where all energies and matrix elements refer to those of the unperturbed integrable Hamiltonian $\hat H_0$. We intentionally separated diagonal and off-diagonal contributions in the expression above. 

It is straightforward to check that for integrable $\hat H_0$ the first (diagonal) term is the most divergent one, because the diagonal matrix elements $O_m\equiv O_{mm}$ and $O_l\equiv O_{ll}$ do not have to be close to each other when the energy difference $\omega_{ml}=E_m-E_l$ becomes of the order of the level spacing, in contrast to what happens in generic systems which exhibit eigenstate thermalization. Therefore, this term is singular. The second (off-diagonal) term [in the second line in Eq.~\eqref{eq:mateleper}] can be divergent as well, but generally it has weaker singularities because even in integrable systems the off-diagonal matrix elements of generic operators are exponentially small in the system size~\cite{leblond_mallayya_19}. This second term may play a more prominent role when the unperturbed Hamiltonian is quadratic (as in our disorder-localized systems) because there only a vanishing fraction of off-diagonal matrix elements is nonvanishing so the nonvanishing matrix elements can be large~\cite{khatami_pupillo_13}.

Hence, we find the most singular perturbative correction to the spectral function $|f_O(\omega)|^2 = \overline{|f^O_m(\omega)|^2}$ (the overline stands for the average over eigenstates), with $|f^O_m(\omega)|^2$ defined in Eq.~(4) in the main text, is
\begin{eqnarray}
\label{eq:f_O_pert}
&&|f_{O}(\omega)|^2 - |f_O^{(0)}(\omega)|^2\approx \nonumber \\ &&\qquad  L{\epsilon^2 \over \omega^2}\left[  \overline {\sum_{l} (O_m-O_l)^2 |V_{ml}|^2 \delta(\omega-\omega_{ml})} \right].\label{eq:spefundiv}
\end{eqnarray}
We note that the linear in $\epsilon$ terms vanish because they are linear in the off-diagonal matrix elements $O_{ml}$ and $V_{ml}$, whose average vanishes~\cite{leblond_mallayya_19}.

Interestingly, Eq.~\eqref{eq:f_O_pert} can be viewed as the Fermi golden rule (FGR) type Lorentzian broadening of the $\delta(w)$ part of the spectral function in the integrable limit by the perturbation. The FGR was recently shown to apply to weakly perturbed strongly interacting integrable systems under unitary dynamics in the context of quantum quenches~\cite{mallayya_rigol_19a} and periodic drivings~\cite{mallayya_rigol_19b}. Indeed for a given state $m$, $L O_m^2$ can be viewed as a $\delta(\omega)$ part of the spectral function or the Drude weight~\footnote{We assume that the microcanonical average of $\hat O$ is zero, otherwise it should be subtracted from $O_m$}. Within the FGR this $\delta$-function broadens to a Lorentzian resulting in:
\begin{equation}\label{eq:fgrO}
    |f^O_m(\omega)|^2\approx |f_{m,0}^O(\omega)|^2+ O_m^2{L\over \pi} {\Gamma_m(\omega)\over \omega^2 +\Gamma_m^2(\omega)},
\end{equation}
where $|f_{m,0}^O(\omega)|^2$ is the spectral function in the integrable limit (not including the Drude weight). This shape of the spectral function was recently observed in numerical calculations in a different model~\cite{schonle_20}.

Comparing Eqs.~\eqref{eq:fgrO} and~\eqref{eq:f_O_pert}, we see that they are consistent if we set 
\begin{equation}
\label{eq:Gamma_n}
\Gamma_m(\omega)={\pi \epsilon^2 }\sum_l \left(1-{O_l\over O_m}\right)^2 |V_{ml}|^2 \delta(\omega-\omega_{ml}).
\end{equation}
The rate $\Gamma_m(\omega)$ is nothing but the FGR rate of change of the normalized variance of $\hat O$ under perturbation $\epsilon \hat V \cos(\omega t)$. Indeed, within the FGR
\begin{multline}
    {d\,\delta O^2(t)\over dt}\equiv {d (\langle \hat O^2(t) \rangle-\langle \hat O(t)\rangle^2\over dt}\\
    =\pi\epsilon^2 \sum_l |V_{ml}|^2 (O_m-O_l)^2 \delta(\omega-\omega_{ml}).
\end{multline}
Defining the decay rate through
\[
{d\,\delta O^2(t)\over dt}=\Gamma_m(\omega)\, O_m^2,
\]
we get Eq.~\eqref{eq:Gamma_n}. Let us finally note that, at small frequencies $\omega$, this rate is expected to be independent of $\omega$ and thus it can be replaced with the static rate $\Gamma_m$ obtained in the limit $\omega\to 0$.

Around the maxima in the typical fidelity susceptibilities in Figs.~1 and~4 in the main text, our numerical results in Figs.~2 and~3 in the main text, and in Fig.~\ref{sm_fig:2}, are consistent with the spectral function exhibiting the previously noted $(\epsilon / \omega)^2$ regime. This is a regime in which perturbation theory breaks down at low frequencies, as made apparent in our numerical results by the fact that the low-frequency parts of the spectral functions exhibit a slower than $(\epsilon / \omega)^2$ divergence.

It is important to emphasize that the $(\epsilon / \omega)^2$ regime is in general absent if $\hat H_0$ is quantum chaotic and $\hat V$ does not break any conservation law. In that case, per the eigenstate thermalization hypothesis, the diagonal matrix elements $O_{mm}$ and $O_{ll}$ are exponentially close (in $L$) to each other when $\omega_{ml}$ approaches the level spacing, and the off-diagonal matrix elements $V_{ml}$ are exponentially suppressed, so that the perturbative correction to the spectral function does not diverge when $\omega$ approaches the level spacing. The $(\epsilon / \omega)^2$ regime is also absent if the perturbation $\hat V$ preserves the integrability of the unperturbed Hamiltonian $\hat H_0$. In that case, the matrix elements $V_{ml}$ have an additional exponential suppression with the system size as $\omega_{ml}$ approaches the level spacing, as shown in Fig.~\ref{sm_fig:1} and in Ref.~\cite{pandey_claeys_20}. As a result, the perturbative correction to the spectral function does not diverge at $\omega$ approaches the level spacing.

\end{document}